\documentclass{article}
\usepackage{graphicx}
\usepackage[letterpaper, left=1in, right=1in, bottom=1in, top=0.75in]{geometry}
\usepackage{amsthm,amsfonts}
\newtheorem{definition}{Definition}

\def\arr{\rightarrow}
\def\del{\partial}
\def\R{\mathbb{R}}
\def\calH{{\cal H}}
\def\({\left(} 	\def\){\right)}

\def\al{\alpha}
\def\la{\lambda}
\def\vp{\varphi}
\def\ga{\gamma}
\def\ze{\zeta}
\def\te{\theta}
\def\om{\omega}
\def\Ga{\Gamma}
\def\La{\Lambda}
\usepackage{graphics}
\begin{document}
\title{The generally covariant meaning of space distances}
\author{S.Capozziello$^{1,2,3}$, A.Chiappini$^{4}$,  L.Fatibene$^{4, 5}$, A. Orizzonte$^{4, 5}$
\and
\small $^1$ Department of Physics  "E. Pancini", University of Napoli {\it``Federico II''}\\
via Cinthia, I-80126, Napoli (Italy)
\and
\small $^2$ Istituto Nazionale di Fisica Nucleare (INFN), Sez. di Napoli,\\
via Cinthia 9, I-80126 Napoli (Italy)
\and
\small $^3$ Scuola Superiore Meridionale, \\
 Largo S. Marcellino 10, I-80138 Napoli (Italy)
\and
\small $^4$ Department of Mathematics, University of Torino, \\
\small via Carlo Alberto 10, I-10123 Torino (Italy)
\and 
\small $^5$ Istituto Nazionale di Fisica Nucleare (INFN),  
\small Sez. di Torino (Italy),\\
\small via P. Giuria 1, I-10125 Torino (Italy).
}                     
%
%
%
%
\maketitle

\abstract{
We propose a covariant and geometric framework to introduce space distances as they are used by astronomers.
In particular, we extend the definition of space distances from the one used between events to non-test-bodies with horizons and singularities so that the definition extends through the horizons
and it matches the protocol used to measure them.
The definition we propose can be used in standard General Relativity although it extends directly to Weyl geometries to encompass a number of modified theories, extended theories in particular.\\
     {\small \it Keywords: Extended theories of gravitation, General relativity, Measurement protocols.}\\

} 
\section{Introduction}
\label{intro}
In a generally covariant theory as General Relativity (GR) there are no Dirac-Bergman observables (other than constants); see, for example, \cite{RovelliObs}.

While distances on spacetime computed with the covariant metric $g$ are generally invariant, they are not endowed with a direct physical meaning.
Although the length of a geodesics can be defined for space-like and time-like geodesics, there is no covariant agreement among observers about which geodesic has to be used to define the 
{\it space distance} between two events or their {\it time separation}.
This is a direct consequence of {\it relativity of contemporaneity}, which is where GR takes its name from.
Two different observers will generally define different space-like synchronisation hypersurfaces and consequently they would disagree on which geodesic to use to define space distances, as it happens already in Minkowski spacetime and Special Relativity (SR). 

So, what do astronomers mean when they say the Moon orbits at $380000km$ from the Earth?
What do cosmologists mean when they say the universe is $14By$ old?

One answer to these questions is that it does not really matter since whatever (reasonable) choice we made, the difference would be below observation precision, anyway.
Another answer is describing a specific observer and its conventions to obtain such a determination, possibly so that the definition reflects the measurement protocols used for it 
and it is at least accurate since no observation can be absolutely precise. 

The second strategy is well established, it is usually based on ADM splittings of spacetime; see \cite{ADM}, \cite{ADMNostro}. 
If one accepts that, only details about how the observer defines the ADM foliation matters.
Such ADM foliations, depending on the situation, can be defined by (various) synchronization conventions 
or it can be driven by symmetries of the spacetime.

In this paper we argue that this is not exactly what it is meant when we determine the distance between the Moon and the Earth.
Of course, it is {\it approximately} what we mean, although not quite.

We think that, as GR claims to be a fundamental theory, it must provide tools to bootstrap physical observations in principle without relying on approximations.
Accordingly, we investigate a fundamental definition which precisely accounts for what astronomers mean by {\it distances}.
This discussion has also a potential impact on definition of deflection angle used in gravitational lensing which, however, we shall not discuss here, since lensing is out of the scope of this paper.  We shall briefly comment in perspectives and leave it to further investigation. 

A good definition of space distances is important in our understanding of a covariant theory even when it is more general than standard GR.
In Extended Theories of Gravitation (ETG) (as well as Palatini formulation) geometry of spacetime is generically a Weyl geometry, i.e.~given by a Lorentzian metric $g$ (or a conformal structure $[g]$) to define pointwise causal structure and a connection $\tilde\Ga$ which determines free fall of test particles (while light rays are shared by the two structures); see \cite{EPS}, \cite{ReviewNostro}.
More generally, often in modified gravity models the geometry of space time is richer than in standard GR. It may have more than one metric of extra scalar fields which, in principle, allows one to define more (conformal) metrics.

To be true this, in principle, happens also in standard GR. If one considers non-vacuum Einstein equations
\begin{equation}
R_{\mu\nu} -\frac{1}{2} R g_{\mu\nu}= 8\pi G T_{\mu\nu} + \La g_{\mu\nu}
\end{equation}
which can be traced to obtain a sort of master equation as it happens in ETG, namely
\begin{equation}
\left(1-\frac{m}{2}\right)R = 8\pi G T+ m\La
\end{equation}
Accordingly, we see that matter produces a scalar field $\vp(x)=T$ (or a linear combination of it with the cosmological constant $\La$).
Of course, this scalar field $\vp(x)$ can be identically zero (for example, for electromagnetism) or constant (for example, for pure cosmological constant and) for some specific types of matter, although in general it is not.

Once one has a scalar field and a metric $g$, although we concede it is not particular natural or motivated in standard GR, one can always define a new conformal metric $\tilde g= \vp g$. 
Strictly speaking, we could argue that most of the experiments that confirm standard GR predictions are in vacuum solutions where this mechanism cannot apply and, every time matter in considered (galaxies, cosmology, recently some issues have been highlighted also in neutron stars), then  extra sources are needed to fit observations.
Accordingly, one can safely imagine to add this conformal transformation by hand.

In this case of standard GR, we are not meaning this is a serious proposal. However, from a foundational point of view one should take a position about how to proceed here and in other modified theories where the trick is considerably less unnatural.

Once we have two metrics, even if they are conformal to each other, they define different notions of space distances and, in principle, one should take a position about which space distances is associated to what we measure as a space distance.

In the context of ETG, as well as in any context where geometry on spacetime is more general than one single metric structure, one should define space distances which have to be used in observations.

On the other hand, when astronomers claim a planet to be at a certain distance from the Sun, what they {\it do} is reporting the {\it radial coordinate} $r$ of pseudo-spherical coordinates centred in the Sun.
Of course, pseudo-spherical coordinates are not defined in terms of distances, as it happens in Euclidean spaces. They are defined as the coordinate in which the metric components take a specific form. 
This definition is well supported by the Newtonian limit, which though does not extend too near the horizon or the singularity.
It can be also phrased in a covariant way in terms of the area of the sphere $(t=t_0, r=r_0)$, which is  a {\it metric} sphere with its induced metric.
In such a way the definition of space distances used by astronomers can also be expressed as a covariant definition although it is quite far from how it is measured in practice.

\medskip
In Section 2 we review various options of ADM distances and show that they do not agree with what is used in astronomy, astrophysics or cosmology.
In ADM-based space distances one has a systematic bias towards higher values for space distances introduced by horizons. 
Generically (of course, depending on the observer), one experience a contraction of {\it ``rulers''} near the horizon which leads to {\it ``overestimation''} of distances. 
Of course, this bias is small, it may be considered acceptable in practice and in comparison with measurement errors.
In this Section we also present a sketch of what must be done in order to explicit the synchronisation procedure. It is hard work even in relatively simple situation (we consider radial distance in Schwarzschild which is essentially the simplest model and in 1+1 dimension).

In Section 3 we state a definition of space distance and discuss how it agrees with astronomical protocols. It is not a surprise (historically, we mean) that astronomical protocols which connects to Newtonian and pre-Netwonian astronomy somehow prefer Minkowski spacetime (as the homogeneous and Lorentzian counterpart of Newtonian spacetime)
and we argue that, in this case, this is not naive. It is instead physically sound and it captures exactly what one needs for a geometric definition when singularities are on the way.

In conclusions, we shall briefly discuss results and highlight connections with the definition of deflection angles in lensing which also require a better definition to cover at once the standard situations and more odd ones, for examples where asymptotic behaviour deviates from flat geometry so light rays can never really be considered asymptotically free.
  
In the Appendix $A$, we collect the computation in the inner region of a Schwarzschild black hole. This is again because, in principle, some inner distance is part of the distance
one is defining and one cannot call for the fact that the inner region is inaccessible and still use the space distance of an event in the external region from the origin (while of course it would make sense to define the distance from the horizon, even though it would still be biased). Also, it is a good exercise of application of synchronisation procedure in a context which is unfamiliar (as it may happen in modified gravity for more theoretical reasons).

\section{Distances in ADM foliations}

An ADM foliation of a spacetime $M$ is a fibration $t:M\arr \R$ which defines the {\it time $t(x)$} at which an event $x\in M$ occurs (for an {\it ADM observer}).
The fiber $t^{-1}(t_0) =\{x\in M: t(x)= t_0\}\subset M$ is called the {\it space} at time $t_0$. Spaces foliate spacetime and are labelled by $t_0$.
One can choose {\it adapted coordinates} $(t, x^i)$ on $M$ so that $x^i$ are coordinates along the spaces and $t$ labels spaces.

Adapted coordinates are generally local even when the fibration is global. If spaces are all diffeomorphic and $t:M\arr \R$ is a bundles, then it is necessarily trivial (see \cite{Steenrod}), i.e.~$M\simeq \R\times S$
for some 3 dimensional manifold $S$, and consequently $M$ is {\it globally hyperbolic}.
If it is not, we shall here consider $U\subset M$ which is globally hyperbolic.

A Lorentzian metric $g$ on $M$ is said to be {\it compatible} with an ADM foliation iff spaces $S_t\subset M$ are space-like submanifolds.
Vice versa, if we have a Riemannian structure $(M, g)$ and we have to choose an ADM foliation we choose it so that $g$ is compatible.
In this case, we say that the ADM foliation is compatible with the metric (or conformal) structure $(M, g)$ (or $(M, [g])$).

The first attitude (first fix the foliation and then consider compatible metrics) is needed when discussing field equations and Cauchy problem. 
Since there the metric is still unknown, one needs to fix a foliation, write down the Cauchy problem for it, solve it and rebuild a metric $g$ on spacetime, which turns out to be compatible by construction. On one fixed ADM foliation one can obtain only metrics which are compatible with it and then one needs to change the foliation to obtain all possible solutions.

The second attitude (first fix the metric) is used when we are given a solution $(M, g)$ of field equations and we want to exploit its geometry. Here we  shall be using the second approach
since we are discussing the meaning of distances {\it in a given solution}. 

Once we have a spacetime $(M, g)$ with a (compatible) ADM foliation fixed on it we can define the {\it spatial distance} between two events, $A$ and $B$,  {\it in the same fiber $S_{t_0}$},
as the minimal length of a geodesic on $(S_{t_0}, \ga )$ connecting the two events, $\ga= i^\ast g$ being the (Euclidean) metric induces on $S_{t_0}$ by the spacetime metric $g$.

A number of (somehow trivial) remarks are in order here.

The metric $\ga$ is Euclidean just because the ADM foliation is compatible with the metric $g$.
Accordingly, the length of geodesics is well defined and positive.

The geodesics on $(S, \ga)$ are not in general geodesics on $(M, g)$ (otherwise, considering $i:S^2\arr \R^3$, geodesics on the sphere, would be geodesics of $\R^3$, i.e.~straight lines, which they are not).

If the two events $A$ and $B$ are not on the same fiber, one needs some extra structure to define their spatial distance.
Essentially, one needs a way of projecting $A$ on the fiber where $B$ is (or vice versa, or projecting both $A$ and $B$ on the same fiber $S_0$).

If we call $(A_0, B_0)$ the projections of $(A, B)$ on $S_0$, the distance between $A_0$ and $B_0$ is called the {\it distance between $A$ and $B$ measured by an observer at $t_0$}. 

In order, to do that we need a congruence of curves $r_x$, transverse to the fibers, hence time-like which are called {\it rest motions}.

Rest motions define another foliation $r:M\arr S$ projecting on an abstract model of space manifold. They also define isomorphisms $r^{t_1}_{t_0}: S_{t_1}\arr S_{t_0}$
for any pair $(t_0, t_1)$.
Accordingly, one can use such isomorphisms to project events on the same fiber and then define the space distance as above. 
Such a distance is even more conventional than the ADM distance is, since it relies on an extra convention, namely the {\it rest motions}.

Let us remark that the tangent lines of rest motions define a connection $\om_r$ on the ADM fibration $t:M\arr \R$, and the rest motions $r$ are integral trajectories for such a connection. 


As long as how one chooses an ADM foliation, also for that the answer is not unique. 
There are essentially two ways which are used in practice. 

The first way is to define some synchronisation convention, e.g.~{\it radar two ways of light synchronisation}.
We pick an observer carrying a clock, i.e.~a parameterised time-like curve $\chi: \R\arr M$, and we consider a light ray going from $\chi$ to $p$ and back to $\chi$.
This singles out two events $p_\pm$ on $\chi$, at which the light rays are emitted and received back, and the corresponding two readings of the clock $s_\pm$.
Then, {\it assuming} that the time to go and to come back are equal, the event $A$ is synchronised to the mid-point $s_p= \frac{1}{2}(s_++s_-)$.
This, in view of EPS axioms (see \cite{EPS}, \cite{Perlick}, \cite{Polistina}, \cite{EPSNostro}, \cite{book2}), defines (in general only locally) an equivalence relation, $p\sim q$ iff $s_p=s_q$, which defines leaves $S_p$ as equivalence classes and, at least locally, an ADM foliation.

This procedure is highly conventional. It relies on the observer choice {\it and} on the assumption that the time spent for the light to go and to come back is the same, and does it not only without testing the claim but {\it before} being able to even define what in principle would be the time spent to go or come back. As the distance between events on different leaves needs extra structure to be defined, the same happens for defining the time lapse between events at a distance.

Besides the synchronisation conventions the time lapse also depends on the choice of a clock, an {\it arbitrary} clock, since in general we  did not say what a {\it uniform} or a {\it proper} clock is, yet.

However, synchronisation is not the only way to specify an ADM foliation. Sometimes, if the isometry group of $(M, g)$ has enough structure, then we can use Killing distribution to define a different foliation, which in fact is not defined using clocks.
This second approach is used in astrophysics and cosmology as well as whenever one cannot send a message to the object and wait for the message to come back, as it is needed by radar synchronisation protocol.

\subsection{ADM foliations of Schwarzschild metric}

A good example is Schwarzschild spacetime $(M, g)$ where one can use both the ADM approaches described above.
Let us here discuss different foliations in a Schwarzschild spacetime. In the next Subsection, we shall discuss distances induced by them.

In dimension $m=4$, being Schwarzschild a static, spherically symmetric spacetime, it allows a Killing algebra which defines at each point generators for 3 rotations and one time translation.
The three rotations define a distribution of rank 2, adding time translations a distribution of rank 3.
The complete Killing distribution of rank 3 defines hypersurfaces $(r=r_0)$ which are cylinders.

The timelike Killing vector alone defines an integrable distribution which produce rest motions $(t=s, x=x_0)$ as integral curves.
One can show there are adapted coordinates $(t, r, \te, \phi)$ in which the time-like Killing vector is in the form $\xi=\del_0$ and
the rotation Killing vectors $\ze_i$ are tangent to the spatial spheres $(t=t_0, r=r_0)$.
In these coordinates, the metric reads as 
\begin{equation}
g= -A(r) dt^2 + B(r) dr^2 + r^2 (d\te^2+ \sin^2(\te) d\phi^2)
\end{equation}
For Schwarzschild solution, one has $A= B^{-1}= 1 -\frac{\al}{r}$ for some positive $\al>0$ which is called the {\it Schwarzschild radius}.
We have two regions. 
In the {\it external region} $(r>\al)$, in which $A>0$, the $t$ direction is time-like, $(r, \te, \phi)$ are space-like directions.
In the {\it internal region} $(r<\al)$, in which $A<0$, the $r$ direction is time-like, $(t, \te, \phi)$ are space-like directions.
On the hypersurface $r=\al$ (on which coordinates are degenerate) there is an apparent singulariry (since $A=0$) and it is called the {\it horizon}.

For the external Schwarzschild metric, the time-like Killing vector $\xi$ is integrable, i.e.~its normal distribution is integrable,
and it defines the ADM space-like foliation $(t=t_0)$.

In the internal Schwarzschild metric, $\xi$ is space-like as $\ze_i$ and accordingly, it still defines an orthogonal foliation, however, leaves $(t=t_0)$ are not space-like, they are Lorentzian manifolds themselves. 
The Killing distribution in the internal region is of rank 3 and it defines a foliation $(r=r_0)$.

Hence the Killing algebra defines two ADM foliations, one $(t=t_0)$ in the external region one $(r=r_0)$ in the internal region.

In order to define ADM foliations, one can alternatively keep stuck to synchronisation protocols. These ADM foliations are less geometrical, although tightly bound to physical measurements protocols. 
A detailed description of synchronisation procedure as a matter of fact tells us how to define protocols for distances.

Let us see how to define some synchronisation surfaces on the Schwarzschild spacetime, either in the external region or in the internal region.
Let us restrict to radial direction so that we can restrict to 2 dimensional Schwarzschild metric 
\begin{equation}
g= -A dt^2 +\frac{dr^2}{A}
\qquad\qquad\qquad\qquad
A(r):= 1-\frac{\al}{r}
\end{equation}

A  radially freely falling observer on the equatorial plane will follow a timelike geodesic, i.e.~a solution of the Lagrangian
\begin{equation}
\bar L=mc\sqrt{ A (t')^2 - \frac{(r')^2}{A} }\> ds
\end{equation}
that is invariant with respect to parameterization, thus $s$ is any parameter and primes denote derivations with respect to the $s$ parameter.

We can collect $(t')^2$ to obtain the Lagrangian
\begin{equation}
\bar L=mc\sqrt{ A - \frac{\dot r^2}{A}}\> dt
\end{equation}
where we selected $t$ as parameter and dots denote derivatives with respect to $t$.
The parameter $t$ is also called the {\it relative time}.
This Lagrangian has a first integral
\begin{equation}
\calH = \frac{\del L}{ \del \dot r} \dot r + \frac{\del L}{ \del \dot \te} \dot \te -L= -\frac{mcA}{ \sqrt{ A - \frac{\dot r^2}{A}} }
\end{equation}
and one can easily solve for the velocities
\begin{equation}
\( \frac{d r}{d t}\)^2= \Phi(r; E)
\qquad 
\( \frac{d r}{d \te}\)^2=  \Psi(r; E)
\end{equation}
where $E$ is the value of the first integral  $\calH$ on a solution.
Since $E$ is constant along a solution it can be computed at initial conditions. 
For example, for an observer which starts at rest at the event $(t=t_0=0, r=r_0)$ one has
\begin{equation}
E= -\frac{mc \sqrt{ r_0-\al} }{ \sqrt{r_0} }
\qquad\iff
r_0 = \frac{m^2 c^2 \al }{m^2 c^2 -E^2 }
\end{equation}

Either ways, we obtain the observer motion by simple integration
\begin{equation}
t(r)= \pm\int_{r_0}^r \frac{dr }{ \sqrt{\Phi(r; r_0)}}
\end{equation}
If we give up to invert for $r(t)$, we already have the motion parameterised in terms of the parameter $r$, namely
\begin{equation}
\ga : \R\arr M: r\mapsto (r, c t(r))
\label{ObFree}
\end{equation}
The sign determines the branch, escaping or approaching the central mass.

Of course, the parameter $r$ is not the proper time $\tau$. However, one has
\begin{equation}
d \tau^2 = \(A -\frac{ \dot r^2}{ A}\) dt^2 =  \(A -\frac{ \Phi}{ A}\) \frac{1}{\Phi} dr^2
\end{equation}
from which one obtains again by a simple integration
\begin{equation}
\tau(r)=\tau_0 \pm \int_{r_0}^r  \sqrt{\frac{ \Phi - A^2}{ \Phi A} } dr
\end{equation}
which is the proper time parameterised in term of $r$ again.

The two integrals are known analytically for Schwarzschild, therefore give me $r$ and one can compute the value of $t(r)$ at which the observer is crossing $r$ and the reading of its proper clock $\tau(r)$.

Here we see why starting from an invariant Lagrangian and using this parameterization is better that using directly the parameterisation with proper time.
The function $\tau(r)$ is a relatively complicated function, which can be proven to be locally invertible, although obtaining an analytical expression for the inverse is quite difficult.
If we fixed the parameterization to be the proper time in the beginning,  now we would be trying to obtain an analytical expression for the function
\begin{equation}
\chi: \tau\mapsto \(t(r(\tau)), r(\tau)\)
\end{equation}
which manifestly involve the inverse of $\tau(r)$, instead obtaining the two analytical (and somehow elementary) functions $\(t(r), \tau(r)\)$.
Even if any observer has its own proper time, it is not necessarily a good idea to used it as a parameter.  

Moreover, proper time is not well defined for light rays (which are null curves). 
The techniques used here will be later used to describe light signals as well.

Alternatively, in the external region $r>\al$, one can fix an observer, e.g.~a parameterized curve 
\begin{equation}
\chi: \R\arr M: s\mapsto \(\frac{s}{\sqrt{A(r_0)}}, r_0\)
\label{ObInf}
\end{equation}
That is a clock, it is normalised so it measures proper time of the observer, it is not freely falling since the trajectory $(r=r_0)$ is not a geodesic,
if not in the weak sense that it is {\it approximately} geodesics if $r_0$ is big enough with respect to $\al$.
These observes can also be regarded as freely falling observers at space infinity.

The sense to understand the limit procedure is not very clear. 
What we mean is that when $r_0$ is big, the observers (\ref{ObFree}) and (\ref{ObInf}) are closed for an interval around the event $(0, r_0)$.

The bigger $r_0$, the bigger the interval in which these observers stay closed and they are a good mutual approximation.
Notice that there $r_0$ is dimensional thus big and small has no absolute meaning and they do not mean much really.
One should say, for example, that they are big with respect of the distance between sending and receiving signals.

\begin{figure}[htbp] 
 \centering
 \includegraphics[width=5cm]{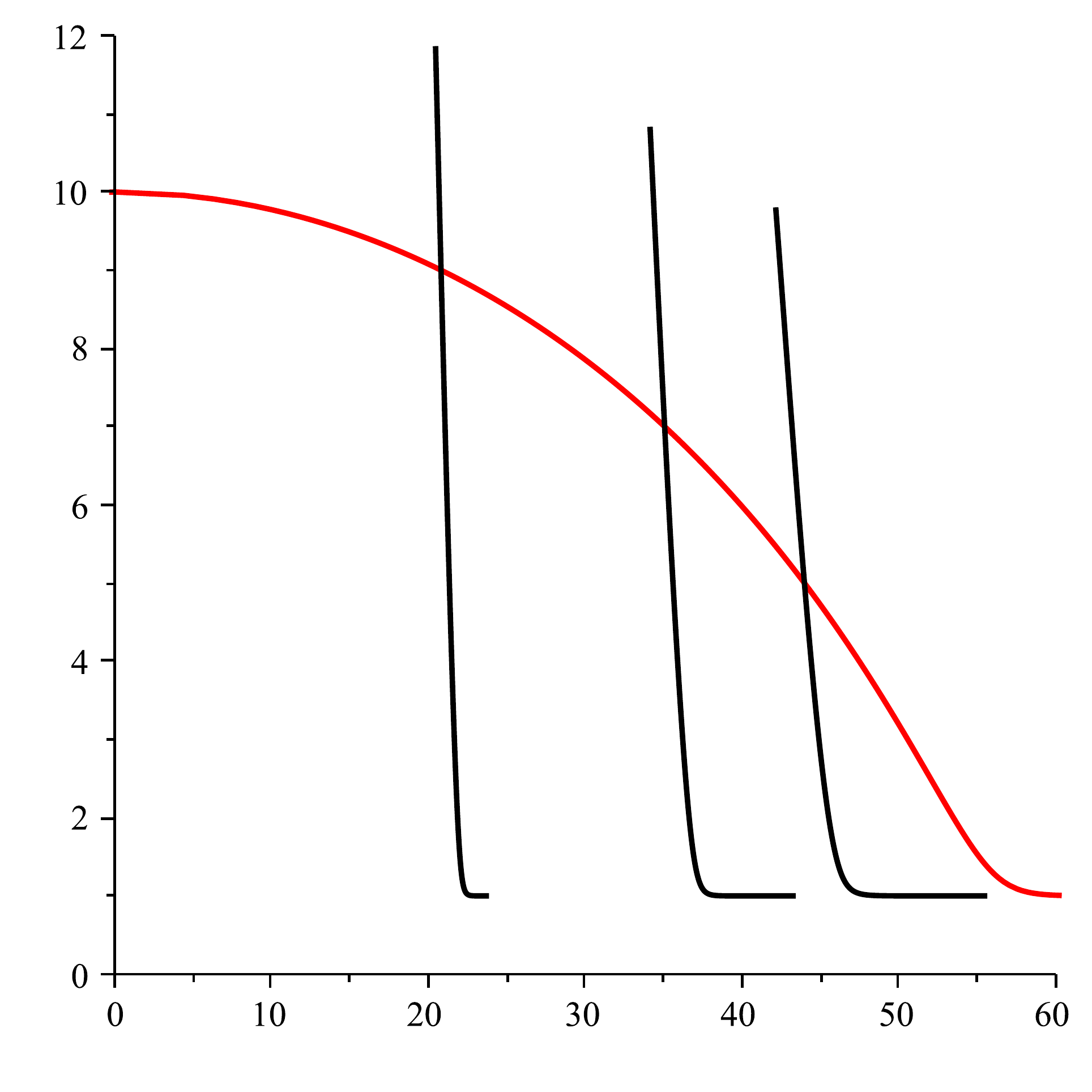}
\qquad
\includegraphics[width=5cm]{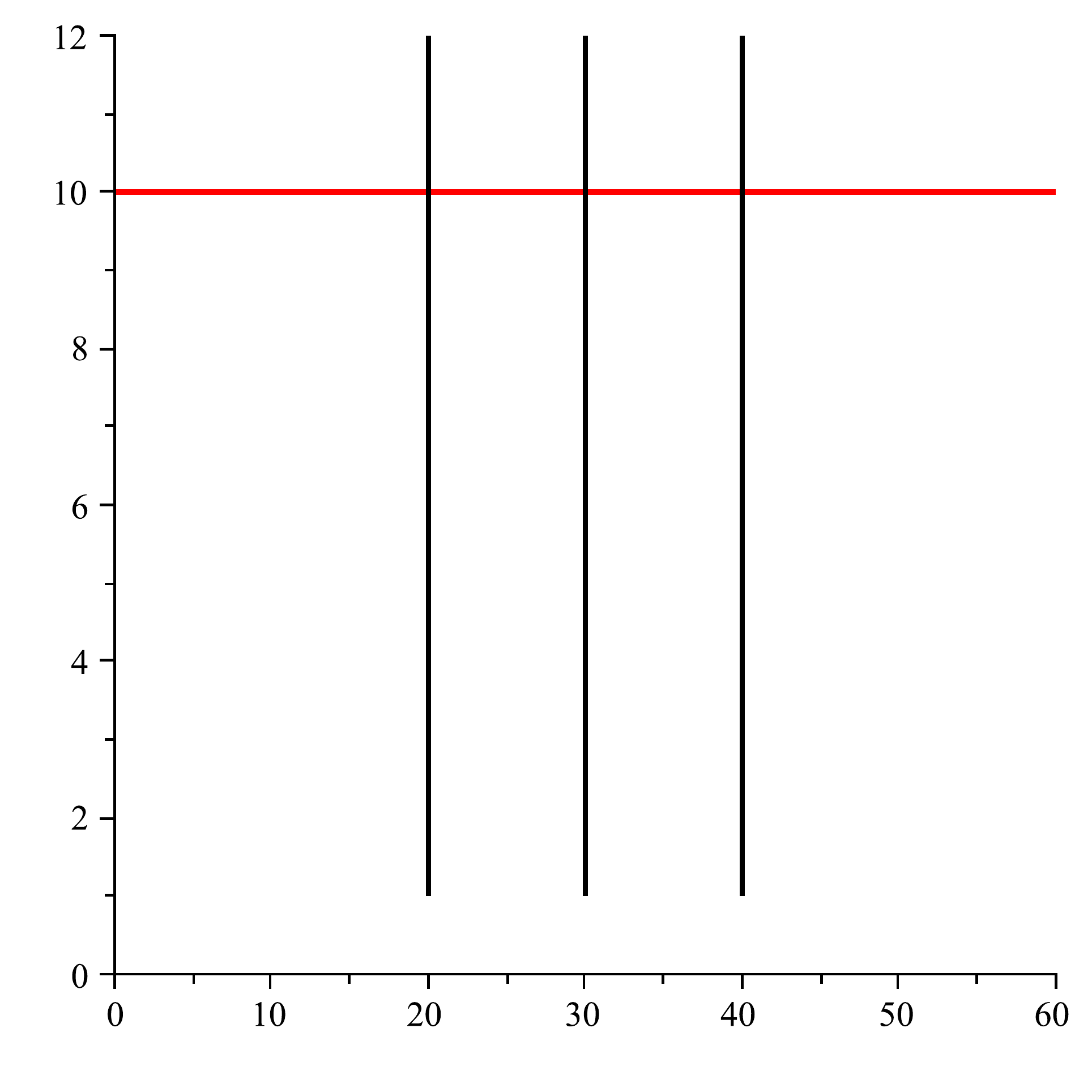}
 \caption{Space manifolds (black) for inertial (\ref{ObFree}) and non-inertial (\ref{ObInf}) observers (red). In both cases, on the axes are $(t, r)$, respectively.}
  \label{fig:1}
\end{figure}

Either ways, we have an observer $\chi$, i.e.~a material point going around with its own proper clock.
In both cases, with such an observer we can define a local ADM foliation.
Let us, in fact, consider an event $p= (t_p, r_p)\in M$ which we want to be synchronised to $\chi(s_0)$. 
That means that one has a message leaving the observer at $s_-=\tau_0 - s$, it is reflected at $p$, and it is received back by the observer at $s_+= \tau_0+s$, so that $s_p =\tau_0$.

In our spacetime we know that light rays are described by the equation
\begin{equation}
t= t_0 \pm \( (r-r_0) + \al \ln \( \frac{ r-\al }{ r_0-\al} \)\)
\end{equation} 
as described in \cite{book2}, \cite{GPS}, \cite{Synge}, \cite{Runge}.

Accordingly, it is easy, in principle, to fix $s$, determining the two light rays $\la_\pm$ passing through $p_\pm:=\chi(\tau_0\pm s)$,
and then determining their intersection $p$ which, by construction, is synchronized to $\tau_0$.
Letting now varying $s$, we get a parametric description of the space manifold, which in this 2d case is a curve, which for example can be explicitly found in the exterior Schwarzschild spacetime.
  
 To be precise we need to stress that we have a {\it local} description of the space submanifolds. 
 If $s$ grows enough light rays may fail to intersect to define points $p$. That is what happens in the internal region.
 Also in the external region if you are too far away, the emission and receiving of the signal will happen too closed to the horizon and fail, for example because the light ray falls into the singularity at $r=0$ before reaching the observer.

For the inertial observer, the space submanifolds are curvilinear space-like and their length defines space distances.
For the non-inertial observer, space manifolds are simply $t=t_0$ as shown in Figure 1.

The situation in the internal region is similar although even more awkward and it is discussed in the Appendix A below.
It is a good exercise to check how the physical intuition (that we have and, unfortunately, it is still firmly Newtonian) can and need to be bended in order to account for covariant geometry with a non-covariant formalism in a situation we are not used to.

\subsection{ADM distances in a Schwarzschild spacetime}

In a Schwarzschild spacetime we have a distance between events, one associated to each ADM splittings we defined above.
If the two events are both internal or both external we simply consider the length of the radial geodesic joining them on a leaf.

Also with a non-freely falling observer, which produces the same foliation of the Killing algebra, the metric induced on the leaf is
\begin{equation}
g = \frac{ dr^2}{ A(r)}
\end{equation}
and the distance of an event $p=(0, r_1)$ from $q=(0, r_0)$ is
\begin{equation}
d_e = \int_{r_0}^{r_1}  \frac{ \sqrt{r} dr}{\sqrt{ r-\al}}
\end{equation}

For example, if we set $r_1=5\> \al$ and we integrate to limit $r_0\arr \al^+$ we get $d_e= 5.916\> \al$.
That is already bigger than $r_1-r_0=4\> \al$ which is the {\it coordinate ``distance''}.

On the other hand, if we set $r_1= 104\>\al$ and $r_0=100\>\al$ we get $d_e=4.020\> \al$, showing that the geometric distance is a good approximation of coordinate distance if we stay away from the horizon, as expected.

If we decide to use the ADM foliation from synchronisation we can repeat the computation.
We can easily instruct Maple or Mathematica to fix an observer, $r_0$ on it which determines a clock reading $\tau_0=\tau(r_0)$, and, for each $s$, compute $\tau_\pm= \tau_0\pm s$, solve $\tau(r)= \tau_\pm$ for $r_\pm$ and the corresponding times $t_\pm=t(r_\pm)$. Then find light rays through the events $p_\pm= (r_\pm, t_\pm)$ on the observer,
choose the signs carefully, and find the intersection of the two light rays, namely $(r(s), t(s))$, which is synchronised with $\tau_0$.

This is enough to draw Figure 1.a, not enough to compute the length of the black line there.

For it, we have to compute the integral
\begin{equation}
d_e^{(i)} = \int_{s_0}^{s_1} \sqrt{\frac{(r')^2}{A}-A (t')^2}\>  ds
\end{equation}
for which we need to know the derivative of $r(s)$ and $t(s)$ with respect to $s$, which means we have to evaluate the displacement of $(r(s), t(s))$ caused by a displacement $s\mapsto s+ds$ at first order. That is easy tracing back the computation described above. 

As a result, if we fix the surface which corresponds to $\tau(5\al)$, from $r_1=6\al$ to $r_0\arr 2\>\al^+$ 
we get a distance  $d_e^{(i)} = 7.026\>\al$.

In the same conditions, but on the surface which corresponds to $\tau(9\al)$, we get a distance  $d_e^{(i)} = 6.173\>\al$.

We see that both do not reproduce coordinate distances near the horizon, although the closer we stay to the $t=0$ axis, the better.

Until now we (fail to) account for external distance, i.e.~distance from the horizon.
One still needs to add the distance between the horizon and the center singularity, which uses the internal solution.
For that we refer to comments in the Appendix A.

 In any event, we clearly see that no reasonable geometrical ADM distance corresponds exactly to $|r_1-r_0|$.
 In other words, while the coordinates $r$ actually parameterizes points in (external) space, it cannot be endowed with a direct physical meaning of distances.
 As a matter of fact, the submanifold $r=r_0$ in Schwarzschild spacetime are {\it metrical} spheres (in space, or cylinders in spacetime).
However, the parameter $r$ is not related to the distance of the surface from the origin. It is instead related to the area of the sphere ($A= 4\pi r^2$ as we see from the angular part of the metric we are here understanding).
 
 This will play a crucial role in the next Section, since $r$ is the radius, i.e.~the distance from the origin, the sphere {\it would have} if it were embedded isometrically into an Euclidean space. As a matter of fact we are going to propose a different way of phrasing this very same claim. 

Of course, for Earth in this situation, using the radial coordinate $r$ as a physical distance would be a very good approximation anyway, but the approximation becomes inaccurate is the get closer to the horizon. 

Therefore we need a way to geometrically interpret what astronomers call a distance, which is not the geometric distance from the central mass in any proper sense, at least not precisely.

\section{Space distances in Schwarzschild spacetime}
 
In view of the discussion above we need to give a geometric meaning to distances reported by astronomers, i.e.~the radial coordinate $r$.
This is not an option: in relativistic theories there is not really separation between geometry and physics.
Endowing physical quantities with a geometric meaning is not a matter of elegance, it is the core of relativity principle.

Also in view of the discussion above, distances reported by astronomers are definitely not geometric space distances based on synchronisation or Killing symmetries.
They seem to be more related to the particular choice of the coordinates $(t, r)$, which is, of course, unacceptable from a geometric viewpoint.
However, this does not rule out a possible geometric principle that {\it in the specific example under consideration} (i.e.~with radial time-like geodesics of a Schwarzschild metric)
could justify space distances which agree with those specific coordinates, though being covariant. 

This would not break general covariance.
What we mean is that there is no {\it a priori} preferred coordinate system.
This has nothing to do with the possibility that {\it in a given solution (i.e.~{\it a posteriori})} there is a preferred coordinate system.

It should not being surprising that, given a FRW solution in cosmology, there are preferred observers which are at rest with respect to the CMBR and this is not breaking general covariance, Lorentz invariance nor in fact Galilei invariance. 

The same happens in presence of a central mass, which of course defines (again {\it a posteriori}) preferred observers {\it at rest}(!) with respect to it.
Being the class of preferred observers determined {\it a posteriori}, this does not break any covariance principle, not the general covariance principle, in particular.

It is simply a gauge fixing one eventually does. As in gauge theories, the later, the better.
 
Here, in fact, we will present a geometric (intrinsic) quantity which, in the specific situation, reduces to $r$ which is (in the specific situation, not as a general principle)
endowed with the physical meaning of space distance from the central mass.

As a matter of fact, when we write down a Schwarzschild metric, we are in fact writing down a whole family of metrics, one for each value of the parameter $\al$ which is a 
measure of the central mass. 
We mean here that each metric ${}^{(\al)}g$ is given on the same manifold $M$ (which is what one understands using one  single coordinate system to write all those metric down).

We showed that in general geometric distance on space submanifolds does not exactly agree with coordinate $r$, with the only (notable) exception of the case $\al=0$, where
the metric ${}^{(0)}g$ becomes Minkowski and the geometric distance on the spatial leaves agrees with the coordinate $r$.
Accordingly, our general proposal, rendered in a precise way, is that astronomer are measuring and giving us the space distance that an event {\it would have} if there were no central mass. That distance is not singular, in the specific situation, agrees with the coordinate $r$, and it can be used as an intrinsic prescription to define space distances in general.

Let us set the following:
\begin{definition}
the {\emph astronomical distance} to be the geometric distance measured by an observer at rest with the central mass, in the limit in which the central mass goes away.
\end{definition}

Of course, this definition is covariant and intrinsic and, although it requires we clarify what it means to set the {\it central mass} to zero, it applies to different formulations of alternative and modified theories of gravity.

In some cases, it is not very clear which parameter corresponds to the central mass but in many cases (and certainly in Schwarzschild) that is pretty clear.

Let us remark how this procedure is similar to what one does in some settings for lensing.

In Eddington-like experiment, one exactly compares two pictures of the sky, one with and one without the Sun along the line of sight.
Stars which are far away from the line through the central mass are used to fix superposition of the two pictures and we measure deviations for the nearby stars. In practice we are measuring the deviation
of the position of a star with respect to the position it {\it should have} if the central mass were not interposed.

In strong, weak, micro, or lensing, we also compare characteristics of the observed stars, with characteristics which are expected (on the base of different assumptions)
if no lens were interposed.

This allows us to measure geometric quantities without relying on assumptions of asymptotic behaviour, not assuming background geometry (or another way of saying the same thing is that we are assuming the geometry with no central mass as a reference background geometry).

\section{Conclusions and perspectives}
 
 In each theory of gravitation we need to discuss what it means to set the central mass to zero.
 For example, in conformal gravity one has BH-like solutions depending on 3 parameters, one simulating cosmological constant (which is clearly not {\it central mass}).
 It is not very clear which combination of the other two parameters needs to be set to zero in order to define the solution with no central mass; see \cite{Marta}, \cite{Misner}.
 
 There are a number of options based on different principles, Newtonian limit, Noetherian conserved quantities, removing the central singularity. 
 
 In the definition of space distance we propose we have a covariant prescription which, once we understand what is the central mass.
 It then defines the astronomical distance and different protocols can be compared and calibrated with it. Being it depending on light rays and time-like observers, it is somehow insensible to the action principle of the model, while it depends on the assumption your model does about how to describe the light motion and free fall. 
Besides that, the prescription is generic enough to adapt to different models.

In our prescription one has a geometry which describe gravity which can be homotopically related to another geometry, which is selected as a {\it reference background} and astronomical distances are defined geometrically in the reference geometry. This is not a general prescription, for example proper time is still defined within the geometry describing gravity. But it is suitable whenever one wants to measure the deviation of one situation from a reference one, as it has been done with covariant conservation laws; see \cite{Augmented}.
The prescription anyway captures the use of space distances in astronomy. As a matter of fact, one should not endow these quantities with a mystical meaning they do not have.
As a matter of fact, a relativistic theory is a covariant theory on spacetime and it describes the geometry of spacetime.
What we usually call the {\it physical meaning} of a quantity is often the remnants of the intuition we have of Newtonian space and time, maybe corrected by SR.
However, in GR, even in classical GR, there is no evolution, no time, no space, only spacetime. Even ADM-foliations which allows to describe single spacetime metric as a sequence of space metrics (and other fields) is a {\it convention}. There is no absolute meaning coming with any protocol to measure time or space although it many be useful to eventually break general covariance to obtain familiar dynamical description of physics.

In a sense, we believe it is even beautiful to have a covariant absolute description of physics, even if it is quite loosely related to observations and then explicitly tracing the protocols which allow us to relate to what we measure. The separation of these two aspects is possible precisely because (and to the extend in which) one starts from an absolute description, as standard GR does. 

Future investigations will be devoted to consider lensing deflection angle,  at least covering definitively asymptotic AdS solutions.
That is a starting point for a wider project which is rebuilding metrology from a generally covariant perspective, instead of founding it on the SR approximation.
In this {\it relativistic metrology} one starts from a class of generally covariant theories, with no predetermined knowledge about the physical world, as required by a theory which claims to be a fundamental theory,
and one defines from scratch and axiomatically the basic metrological objects, possibly relating directly to measurement protocols. 
At the very least, this gives an explicit account of the convention and choices one does when measuring something. 
The hope (and the evidence in this simple example) is that this relativistic metrology comes really separated from dynamical aspects of the gravitational theory
(which now is not in GR) and possibly it applies to a wider class of theories sharing some kinematical characteristics.

Just doing that, one is then able to reliably compare different theories on the basis of observations, design experiments and learn something about the real world which is not implicitly based on previous theories (that we know are wrong). That is what we understand a fundamental theory should do.

\section*{Appendix A: Internal geometry}

Here we would like to describe what happens if one wants to define an ADM foliation by synchronization in the inside region as we did in the external one.
It is not difficult to describe a time-like geodesic in the inner region (just remember that $r$-direction is the time-like one, not $t$).

The first integral $E$ is a scalar on $TM$ so one can use it to connect observers in the two regions although they are disconnected by the horizon.

The technique illustrated in the external region based on Weirstrass functions does the trick.
Similarly one can define the synchronisation surface as outside.

\begin{figure}[htbp] 
   \centering
 \includegraphics[width=6cm]{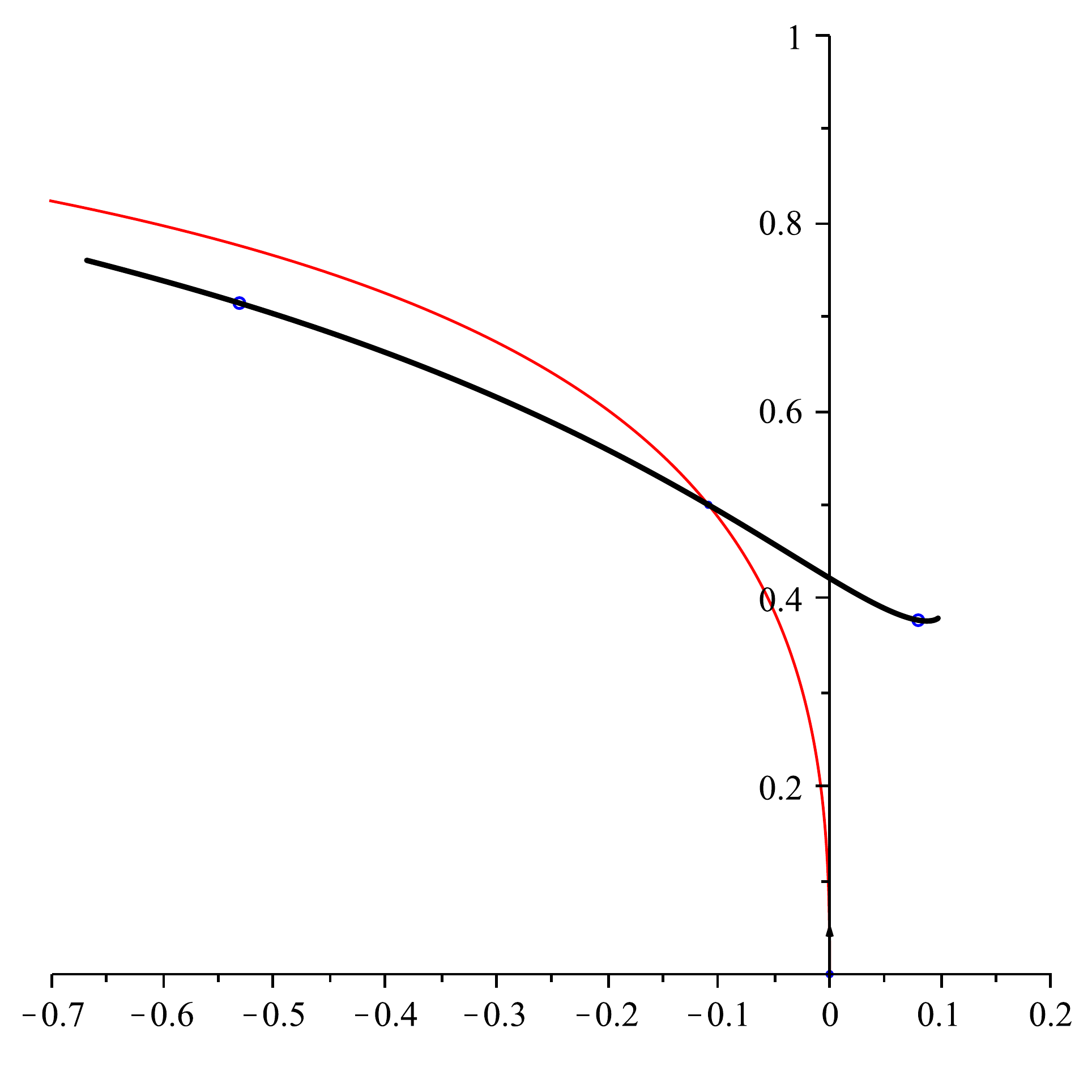}
\qquad  \includegraphics[width=6cm]{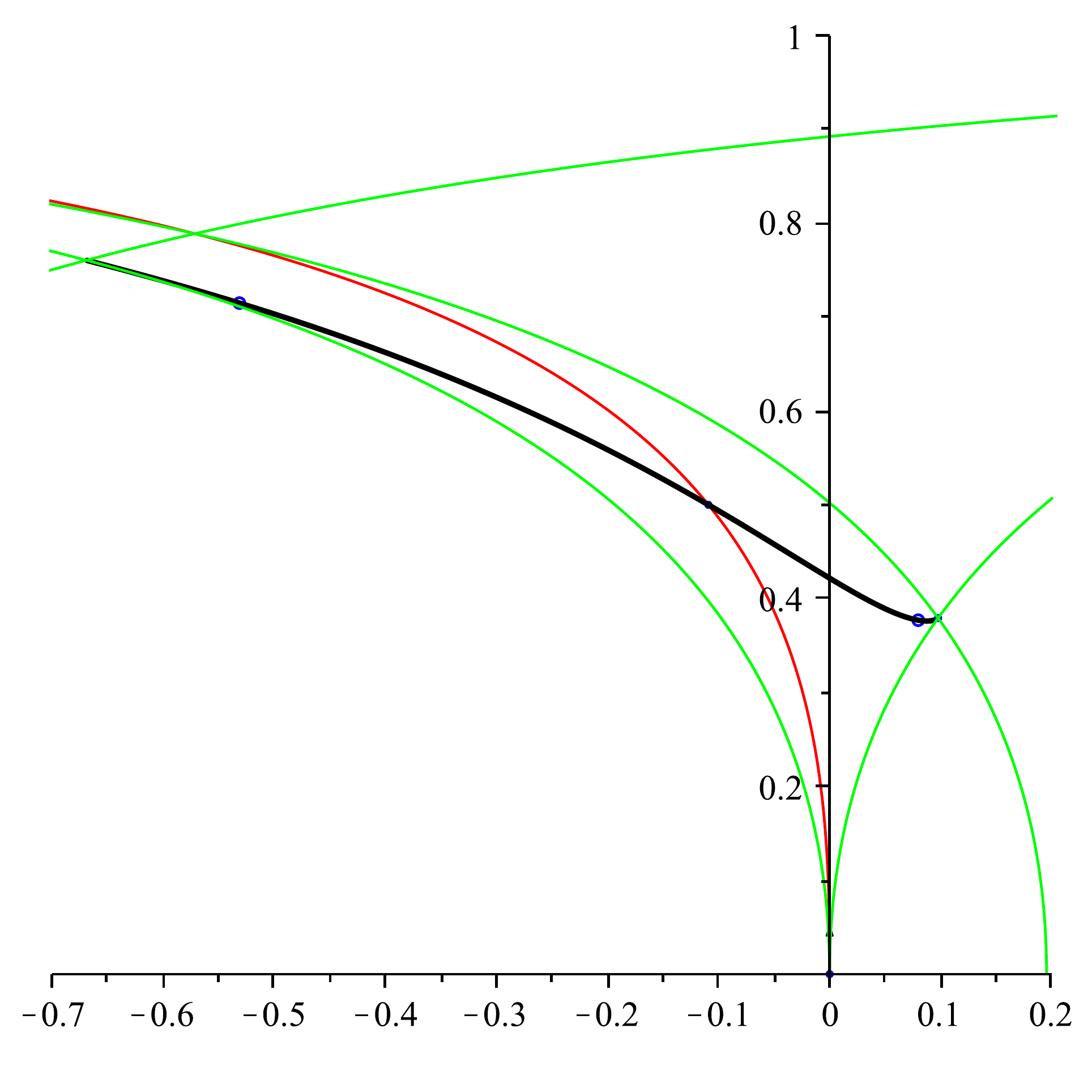}
\caption{a) a freely falling observer in the internal region (red) and a synchronization surface through one of its point (black).
b) a freely falling observer in the internal region (red) and a synchronization surface through one of its point (black). The green lines are light signals exchanged.
We see that if we had to extend the surface further, there will be no time to exchange signals with the observer as the light ray would be captured by the singularity before reacing back the observer.}
   \label{fig:2}
\end{figure}

In the internal solution, we have $A<0$ and $r$ direction becomes time-like while $t$ direction becomes spacetime.
One can characterize freely falling observers which are in falling in a finite proper time.
Each freely fall observer looks for synchronisation surfaces, which in this case are definitely local since if an event is considered which is far enough from the observer it may happen that light signals cannot be exchanged back and forth before the observer (and the light rays itself) falls into the singularity.

Different freely falling observers defines different synchronisation surfaces (even if they are simply translated along $t$ direction) so one cannot find a single synchronisation space submanifold covering  the whole interior region.
This shows how being synchronized is not an equivalence relation. Two events which are synchronized {\it by one observer} are not synchronised for a different observer, even if the observer is simply translated in the internal region.

Locally however, geometric distances do not reproduce $r$-distances.

\begin{figure}[htbp] 
   \centering
\includegraphics[width=6cm]{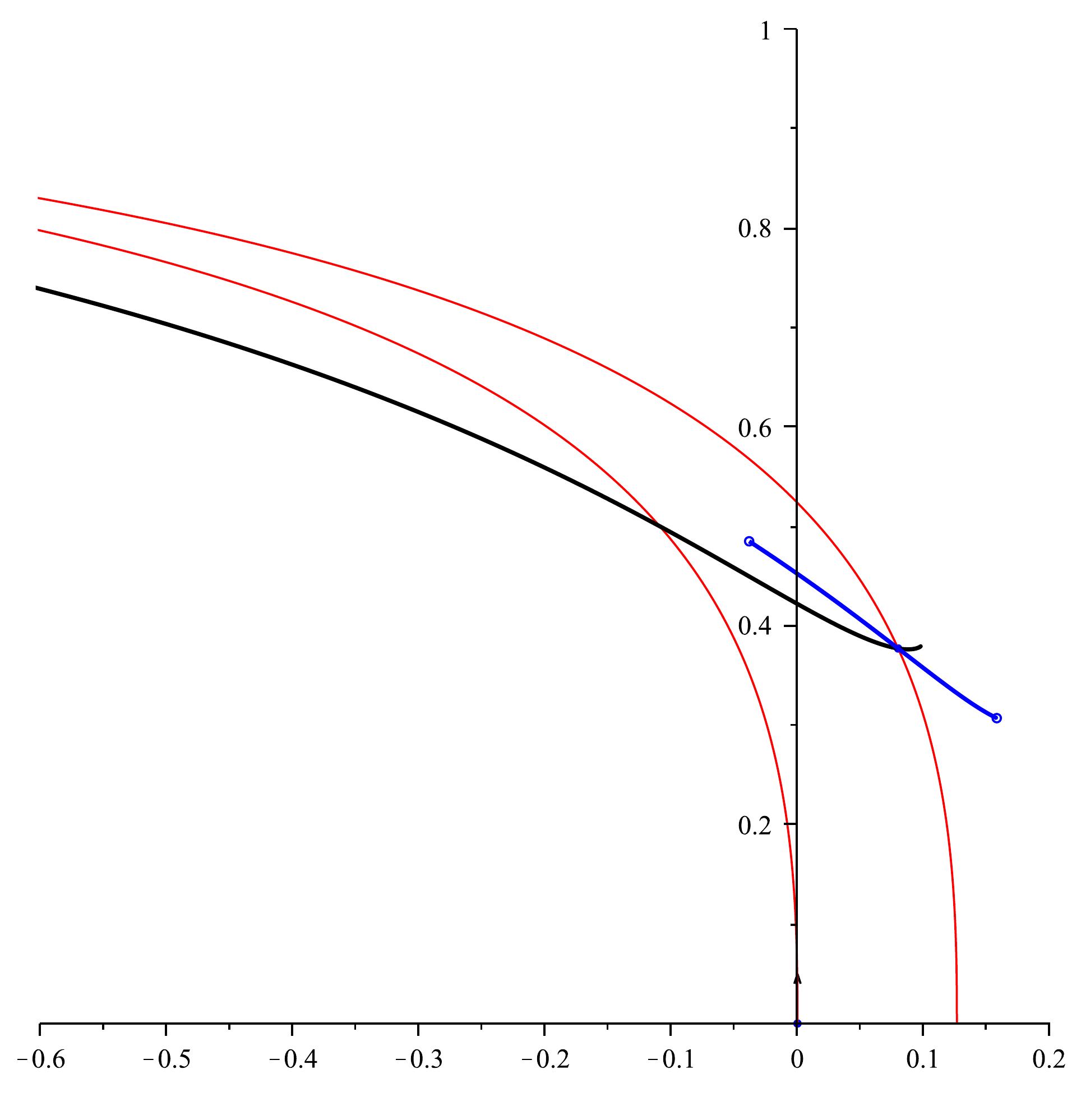}   
\caption{The synchronization surfaces traced by translated observers.}
   \label{fig:3}
\end{figure}

Let us also remark that all synchronization protocol is based on {\it coincidences} (i.e.~intersections of geodesics) and it is hence intrinsic.
As a matter of fact one can find coordinates in which the apparent horizon (at $r=\al$) is not singular and coordinates extends on either sides.
Examples of these coordinate systems are Eddington-Finkelstein (EF) or Kruskal coordinates.
One can repeat the synchronisation procedure in these coordinates, thus having a smooth, non-singular pictures of what happens crossing the horizon.
Also in these coordinates, one cannot extend the synchronization surface down to the real singularity (at $r=0$). This is because the fact that events are not close enough to the observer to exchange messages has nothing to do with coordinates.

\begin{figure}[htbp] 
   \centering
    \includegraphics[width=6cm]{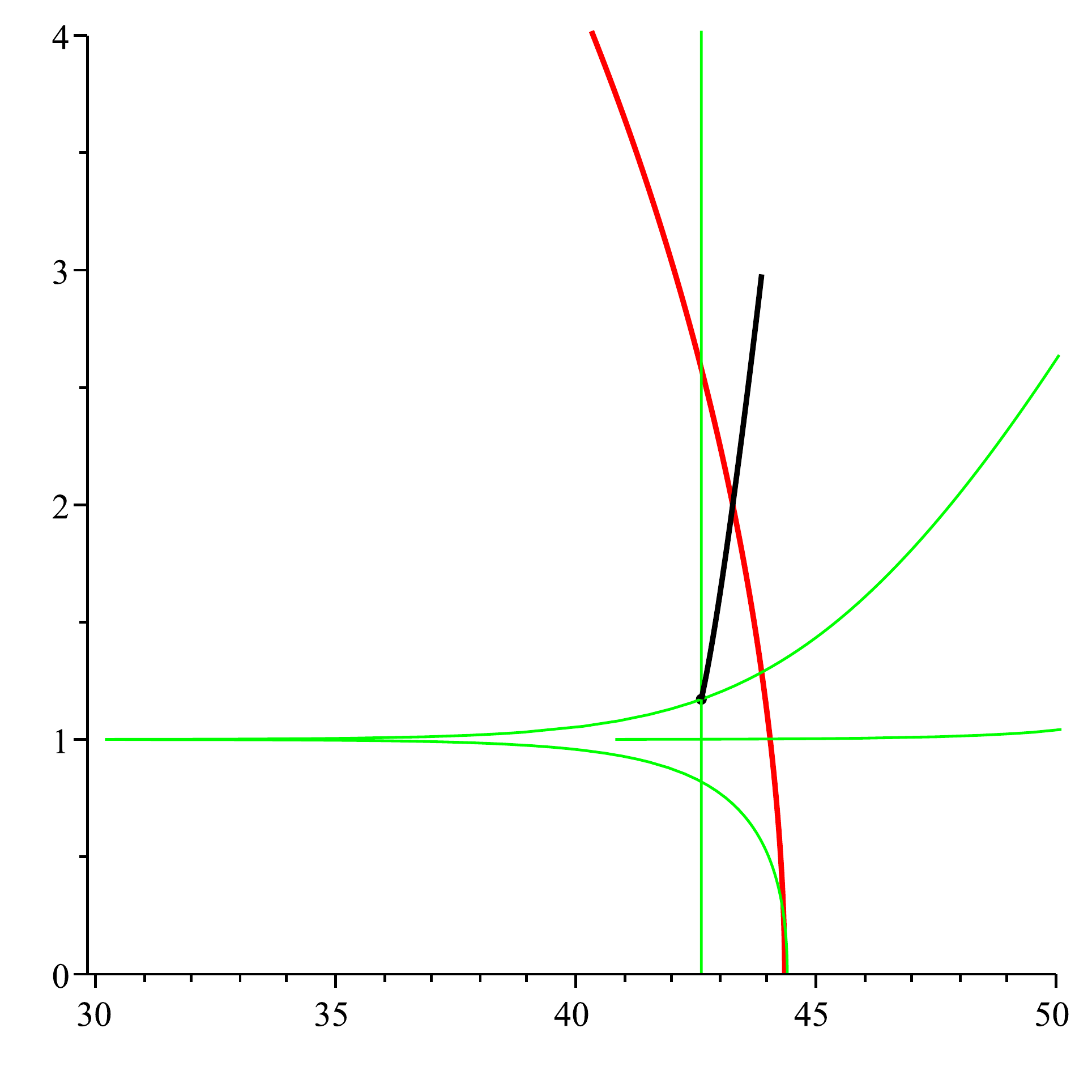}
   \caption{The synchronization surfaces traced in EF coordinates.
We used the same notation of Figure 2.b.
You see that if we try to extend the synchronization line for smaller $r$, below the {\it last light ray from the observer}, the light ray ``shot back to the observer'' falls into the singularity before reaching the observer. }
   \label{fig:4}
\end{figure}

\section*{Acknowledgements}

This article is based upon work from COST Action (CA15117 CANTATA), supported by COST (European Cooperation in Science and Technology).
We acknowledge  the contribution of INFN (IS-QGSKY, IS-MOONLight2 and IS-EUCLID), the local research project {\it Metodi Geometrici in Fisica Matematica e Applicazioni (2019)} of Dipartimento di Matematica of University of Torino (Italy). 
This paper is also supported by INdAM-GNFM.

%
%
%

%
%

\end{document}